\documentclass{ws-procs9x6}

\begin{document}

\title{High-t Meson Photo- and Electroproduction:\\ a Window on the
Partonic Structure of Hadrons}

\author{J.-M. Laget}

\address{CEA-Saclay, DAPNIA/SPhN \\ 
F91191 Gif-Sur-Yvette, France \\ 
E-mail: jlaget@cea.fr}


\maketitle

\abstracts{A consistent description of exclusive photoproduction of mesons 
at high momentum transfer $t$ relies on a few effective degrees of freedom which
can be checked against Lattice calculation or more effective models of QCD.}

\section{Introduction}

Exclusive photoproduction of mesons at high momentum transfer $t$ offers us a
fantastic tool to investigate the partonic structure of hadrons. The high
momentum transfer $t$ implies that the impact parameter is small enough to allow
a short distance interaction between, at least, one constituent of the probe 
and one constituent of the target. In addition, the exclusive nature of the
reaction implies that all the constituents of the probe and the target be in the
small interaction volume, in order to be able to recombine into the well defined
particles emitted in the final state.

This enables us to identify, to determine the role and to access the
interactions between each constituent of hadrons.

The relevant constituents depend on the scale of observation. At low
momentum transfer $t$, a comprehensive description~\cite{gui,jml} of available
data is achieved by the exchange of a few Regge trajectories, between 
the probe and the target. At very high (asymptotic) momentum transfer $t$, the
interaction should reduce to the exchange of the minimal number of gluons, in
order to share the momentum transfer between all the current quarks which
recombine into the initial and final hadrons~\cite{bro}. Here, dimensional
counting rules lead to the famous power law behavior of the various cross
sections, but it is unlikely that the present generation of high luminosity
facilities gives access to this regime.

To date, CEBAF at Jefferson Laboratory is the only facility which allows to
reach  momentum transfers $t$ up to 6 GeV$^2$ in {\em exclusive} reactions, at
reasonably high energy ($s$ up to 12 GeV$^2$). This range will be considerably
enlarged when the 12 GeV CEBAF upgrade becomes a reality.  

The range of momentum transfers accessible at JLab corresponds to a resolving
power of the order of $0.1$ to $0.2$~fm. It is significantly smaller than the
size of a nucleon, but comparable to the correlation lengths of partons (the
distance beyond which a quark or a gluon cannot propagate and hadronizes).  At
this scale, the relevant degrees of freedom are the {\em constituent partons},
whose lifetime is short enough to prevent them to interact and form the
mesons which may be exchanged in the $t$-channel, but is too long to allow to
treat them as current quarks or gluons. This is the {\em non perturbative partonic
regime}, where the amplitude can be computed as a set of few dominant Feynman
diagrams which involve dressed quarks and gluons, effective coupling constants
and quark distributions~\cite{cano}. This resembles the treatment of meson
exchange mechanisms at low energies~\cite{physrep}.

The first results~\cite{phiprl,bat}, recently released by JLab, support such a 
picture and provide a link with Lattice QCD predictions on the gluon 
propagator~\cite{will}, modelisations of quark wave functions in 
hadrons~\cite{kro} and Regge saturating trajectories~\cite{gui}.

\section{Gluon exchange}

	Measurement of phi photoproduction selects gluon
exchanges. Since the $\phi$ meson is predominantly made of a
strange $s\overline{s}$ quark pair, and to the extent that the strangeness
content of the nucleon is small, quark interchange mechanisms are suppressed. 

\begin{figure}[th]
\centerline{\epsfxsize=1.5in\epsfbox{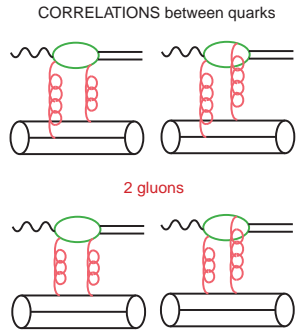}}   
\caption{The four diagrams depicting two-gluon exchange mechanisms.
\label{2gluons}}
\end{figure}

\begin{figure}[th]
\centerline{\epsfxsize=2.0in\epsfbox{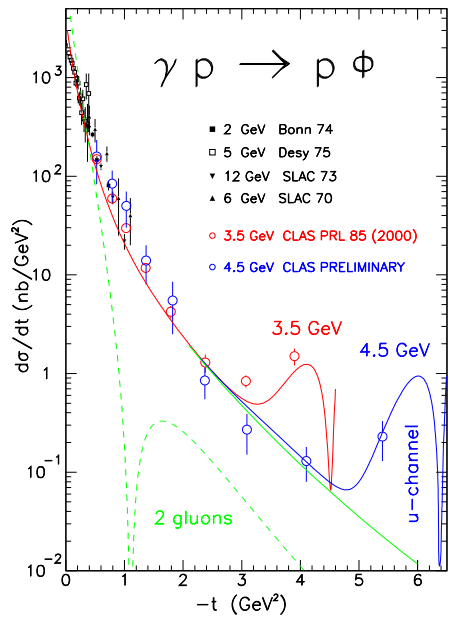}}   
\caption{The $\phi$ meson photoproduction cross section at $<E_{\gamma}>=$ 3.5 
and 4.5 GeV.
\label{phi_cross}}
\end{figure}

The destructive interference between the two graphs in the bottom of 
Fig.~\ref{2gluons} (where each gluon couples to the same quark in the nucleon, 
but may couple to a different
quark in the vector meson) leads to a node in the cross-section (dashed line)
depicted in Fig.~\ref{phi_cross}. This node is filled when the two gluons are
allowed to couple also to two different quarks in the proton (faint solid line),
giving access to their correlation.
At the largest momentum transfer $t$, nucleon exchange in the $u$-channel takes
over, but the corresponding peak move towards higher $t$ when the incoming
photon energy increases (solid curves, marked 3.5 and 4.5 GeV), leaving more
room to access the two gluon exchange contribution.\\

The combined use of the high luminosity of CEBAF and of the large acceptance of
the CLAS set up at JLab made possible the measurement of  cross sections, more
than two orders of magnitude below previous ones, in a virgin domain up to $t$=
5.5~GeV$^2$. The data at $E_{\gamma}$= 3.5 GeV have been published~\cite{phiprl},
while the data at $E_{\gamma}$= 4.5 GeV are still preliminary and their errors
will soon decrease. This is however good enough to exhibit
the trend of the cross section, to confirm the move of the $u$-channel peak 
toward higher momentum transfers and to establish the relevance of the two 
gluon exchange description. 

The key to the success of such a good agreement is the use~\cite{cano} of the 
correlated quark wave function of Ref.~\cite{kro} and the Lattice gluon 
propagator of Ref.~\cite{will}. For instance, if a perturbative gluon propagator
had been used, one would have obtained a much more steep $t$ dependency of the
cross-section (see Ref.~\cite{phiprl} for more details). In the gluon loop
(Fig.~\ref{2gluons}), the virtuality of each gluon is on average $t/4$, {\it
i.e.} about 1~GeV$^2$ at $t=$ 4~GeV$^2$ where, according to Fig.~2 in
Ref.~\cite{will}, the lattice gluon propagator exhibits strong non perturbative
corrections. On the contrary, it reaches it asymptotic behavior around a
virtuality of 4~GeV$^2$: this requires a momentum transfer $t$ of about
16~GeV$^2$, which will be achievable when CEBAF is upgraded to 12 GeV. 

I refer to the talk of K. McCormick~\cite{mcc} for the analysis of the tensor 
polarisation of the $\phi$ which confirms the dominance of the $u$-channel 
contribution at the highest momentum transfer.

\section{Quark exchange}

Quark exchange mechanisms are not supressed in the photoproduction of $\rho$ and
$\omega$ mesons, which are mostly made of light quarks. 

\begin{figure}[th]
\centerline{\epsfxsize=2.0in\epsfbox{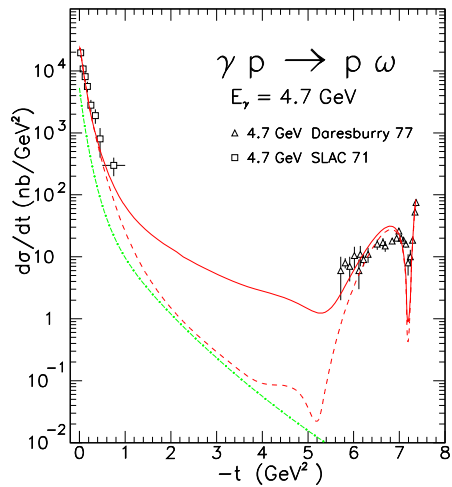}}
\vspace{-0.5cm}   
\caption{The $\omega$ meson photoproduction cross section at $E_{\gamma}=$ 
4.7 GeV.
\label{omega_cross}}
\end{figure}

The most stricking example is the photoproduction of $\omega$ meson
(Fig.~\ref{omega_cross}). At low momentum transfer, $\pi$ exchange (dashed
line) takes over two gluon exchange (dash-dotted line). At the largest momentum transfer,
$u$-channel proton exchange dominates and accounts for the backward angle cross
section. Here the experimental node is well reproduced by the use of a
non-degenerate Regge trajectory for the nucleon. At intermediate transfer, the
use of a linear Regge trajectory for the $\pi$ leads to a vanishing cross
section, while the use the saturated Regge trajectory (full line), which 
already led to a
good accounting of the cross section of the $\gamma p \rightarrow \pi^+ n$
reaction~\cite{gui}, enhances the cross section by two orders of magnitude. 

This is really a parameter free prediction, which has been beautifully confirmed
by the recent CLAS data: I refer to the talk of M. Battaglieri~\cite{bat}, for
a more detailed discussion in the $\omega$ sector as well as the $\rho$ sector. 

\section{Compton scattering}

\begin{figure}[th]
\centerline{\epsfxsize=2.0in\epsfbox{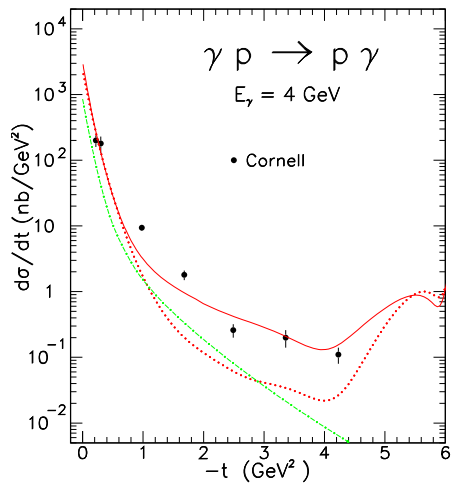}}   
\vspace{-0.5cm}   
\caption{The Real Compton Scattering cross section at $E_{\gamma}=$ 
4.0 GeV. Dash-dotted line: two gluon exchange. Full (dotted) line: saturated 
(linear) Regge trajectories.
\label{compton_cross}}
\end{figure}

A few GeV real photon has a significant hadronic component. Due to the
uncertainty principle, it fluctuates into vector mesons (or quark anti-quark
pairs of various flavors) over a distance which exceeds the nucleon size. For
instance, a 4 GeV real photon fluctuates into a $\rho$ meson over about 2.7~fm.
The real Compton scattering cross section is therefore related to the $\rho$
meson  photoproduction cross section by a simple multiplicative factor: 
$4\pi \alpha_{em}/f_V^2$, where $f_V$ is the radiative decay constant of the 
vector meson. As shown in Fig.~\ref{compton_cross}, the comparison with this
model~\cite{cano} and the old Cornell data confirm this conjecture.

\begin{figure}[th]
\vspace{-1.0cm}   
\centerline{\epsfxsize=2.5in\epsfbox{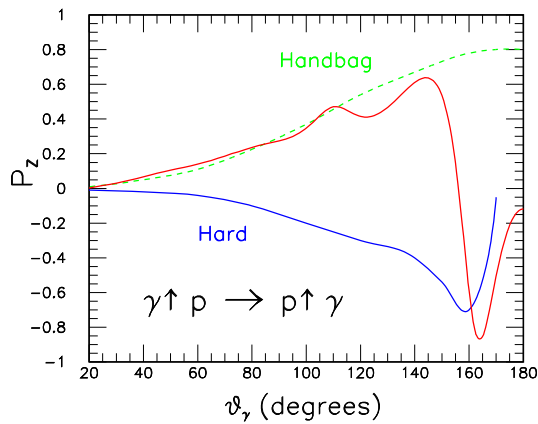}}   
\vspace{-1.5cm}   
\caption{The longitudinal spin transfer coefficient in Real Compton Scattering 
at $E_{\gamma}=$ 4.0 GeV.
\label{compton_pol}}
\end{figure}

More interesting, the model predicts also spin observables. As an
example, Fig.~\ref{compton_pol} compares the predicted longitudinal spin
(helicity) tranfer (solid line in the upper part) to the hard
scattering models~\cite{van} (solid line in the bottom part) and the soft
"handbag" model~\cite{kro2} (dashed line). The strong variation at backward
angles comes from the $u$-channel baryon exchange.
The preliminary data from Hall A at JLab confirms
this prediction. I refer to the talk of A. Nathan~\cite{nat}, for a more
detailed account and discussion of future experimental prospects.

\section{Conclusion}

A consistent picture is emerging from the study of exclusive photoproduction
of vector meson and Compton scattering. At low momentum transfer, it relies on
diffractive scattering of the hadronic contents of the photon, in a wide 
energy range from threshold up to the HERA energy domain. At higher momentum 
transfer, it relies on a partonic description of hard scattering mechanisms 
which provides us with a bridge with Lattice Gauge calculations. The dressed 
gluon and quark propagators have already  been estimated on lattice. One may
expect that correlated constituent quark wave functions (at least their first
moments) will soon be available from lattice. An estimate, on lattice, of the
saturating part of the Regge trajectories is definitely called for. 

To day, JLab is the only laboratory which allows to explore this regime, 
thanks to its high luminosity.
Its current operation, at 4--6 GeV, has already revealed a few
jewels. Its energy upgrade to 12 GeV will expand to higher values the 
accessible range in
momentum transfer and allow to extend these studies to virtual photon
induced reactions. It will permit, among others, a more
comprehensive study of the onset of asymptotic hard scattering regime
and of the role of  correlations between quarks.

\section*{Acknowledgments}
This work has been partly supported by European Commission under Contract
HPRN-CT-2000-00130. Part of the results presented in this note arose from
collaboration and discussions with F. Cano, over the past two years.

\end{document}